# VIRIAL MODELS AND ANISOTROPY OF VELOCITY DISPERSION IN E-GALAXIES

**Kondratyev B. P.**[1,2] , **Kireeva E. N.**[1]

[1]Sternberg Astronomical Institute, M.V. Lomonosov Moscow State University, 13 Universitetskij prospect, 119992, Russia
[2]Central Astronomical Observatory at Pulkovo, Russia

E-mail: work@boris-kondratyev.ru

**ABSTRACT**

A tensor virial approach is used to construct detailed models of 22 flattened ellipticals. The models are combined with the new observational data: extended surface brightness distributions, the profiles of isophotes flattening and their twisting, the rotation curves and the velocity dispersion profiles. A key feature of these models is a rigorous consideration of the influence of the spatial shape of galaxies (oblate spheroid or triaxial ellipsoid) on the dynamics, as well as the structure of the internal density layers in these systems. For each galaxy the ratio of the rotational energy to the potential energy $t = \frac{T_{rot}}{|W|}$ was found. Comparing this ratio with the observed $\frac{v_{rot}}{\sigma}$ it is concluded that the majority of these systems cannot have isotropic velocity dispersion tensors. The anisotropy parameter $\beta$ is limited to the interval $0.0 \leq \beta \leq 0.224$. An including of the gradient $M/L$ profile decreases this parameter by 21% for the M87 galaxy. A correlation was found between β and flattening of the outer regions for giant E-galaxies. Our results are compared with the results of other researchers. It was found that for small fast rotators our values $\beta$ are in good agreement with those, obtained in the $ATLAS^{3D}$ project. However, for giant E-galaxies, our models provide better agreement with observations, than axisymmetric JAM models. In addition, our velocity dispersion anisotropy results are in satisfactory agreement with the results of high resolution cosmological simulations in the Illustris project.

**Key words: galaxies: elliptical and lenticular, E and cD – galaxies: kinematics and dynamics.**

## 1. INTRODUCTION

Until the end of the 70s of the twentieth century it was believed that elliptical galaxies have axial symmetry, and their flattening is due to rotation. However, this conclusion was not based on observations, but on the consequences of the classical theory of equilibrium figures. Taking into account the replacement of an ideal fluid by a collisionless stellar medium, it was assumed that the prototypes of E-galaxies are similar to modified Maclaurin spheroids. Collisionless phase models, based on these considerations (Gott 1975; Wilson 1975 and Larson 1975), had the shape of an oblate spheroid and phase density depending on two integrals of motion: energy $E$ and angular momentum $J$. In models of this type, the condition of pressure isotropy in the meridional plane $\sigma_{33}^2 = \sigma_{12}^2$ was satisfied, as a result of which the model flattening depended on its rotation like in classical Maclaurin spheroids. More complex ellipsoidal phase models were constructed in (Kondratyev 1995).

The situation dramatically escalated once the rotation of the galaxy NGC 4697 was actually measured (Bertola & Capaccioli 1975); it turned out that the magnitude of this rotation is small and obviously insufficient to create the observed flattening. Small rotation was soon confirmed on a wider selection of E-galaxies (Illingworth 1977), and it eventually led to a revision of views on elliptical galaxies. Note two sides of this problem.

Hot on the heels of this discovery, Binney (1978) proposed the idea of residual anisotropy of velocity dispersion in E-galaxies, which fit well into the overall picture of the dynamics of collisionless stellar systems. Pressure anisotropy in E-galaxies was studied in more detail in (Kondratyev 1981), where the real variation of isophote ellipticity were taken into account for the first time.

The problem of the dynamics of E-galaxies is closely related to the question of the spatial shape of these star systems. Any elliptical galaxy is represented by an ellipse (limb) in the projection on the picture plane. However, this ellipse may be a result of projection of not only an oblate spheroid, but also a triaxial ellipsoid, so there are difficulties in interpreting observations: what actual spatial shape does one or another elliptical galaxy have? This question is important, since the dynamic properties of models of oblate and triaxial ellipsoids can vary greatly.

To clarify the question of the shape of E-galaxies, two observational tests were proposed (Kondratyev & Ozernoy 1979), see also (Contopoulos 1956; Stark 1977). The first test indicates that in the case of a triaxial galaxy its axis of rotation most likely will not coincide with the visible minor axis of the projected ellipse (misalignment between galaxy's angular momentum vector and its minor-axis, see Statler 1991). The second test is based on the effect of isophote twisting, often observed in E-galaxies. The use of these tests allowed us to find out that there are both axisymmetric oblate spheroids and triaxial ellipsoids among E-galaxies. Note that the question of the orientation of galaxies with respect to the axis of rotation is wider than for liquid equilibrium figures. In particular, unlike liquid figures, collisionless triaxial ellipsoids (even without internal centroids flow) can rotate stably around not only the minor, but also the middle axis (Kondratyev 1983). In this regard, it should be noted that there are observed galaxies with rotation around the major axis (Davies & Birkinshaw 1986, 1988; Franx, Illingworth & Heckman 1989; Jedrzejewski & Schechter 1989).

According to their morphology, elliptical galaxies can be divided into two subclasses. One subclass includes giant galaxies that often have very slow rotation; it is believed that their triaxial form is largely supported by the anisotropy of velocity dispersion (slow rotators, Emsellem et al. 2004; Cappellari et al. 2007). Another subclass consists of small, regular E-galaxies. They have a form of an oblate spheroid, which is maintained by their rotation, and the internal pressure can be considered almost isotropic (at least, see above, in the meridional plane). Such a classification mostly coincides with the kinematical separation of galaxies into fast and slow rotators, proposed in the SAURON project (Cappellari et al. 2007; Krajnović et al. 2008). Of course, none of these classifications covers the whole variety of early-type galaxies, since the kinematical and morphological features of every star system depend on its individual history of its formation.

A characteristic feature of E-galaxies is that their internal structure is not determined by total mass and angular momentum alone (de Zeeuw & Franx 1991; see also Davies & Illingworth 1983). With the development of new concepts of elliptical galaxies, the problem of construction of dynamic models, in which equilibrium is created not only by gravity and rotation (where the choice of the spatial shape of the star system is important), but also by the anisotropy of the velocity dispersion, has become an issue of the day. With the help of models we should find out how much the velocity ellipsoid in a particular galaxy differs from the sphere. It should be emphasized that it is important to choose an adequate mathematical apparatus in the construction of such models. Our approach is based on the application of the virial tensor theorem to an ellipsoidal layered-inhomogeneous subsystem and includes models with rotational symmetry and without it (Kondratyev 1982, 1989).

To measure the velocity dispersion anisotropy in the system, we can use the value

$$\beta = 1 - \frac{\sigma_{33}^2}{\sigma_{12}^2}, \qquad (1)$$

where $\beta = 0$ corresponds to the case of isotropic pressure. Given the ratio $\sigma_{11} = \sigma_{22} = \frac{\sigma_{12}}{\sqrt{2}}$, the square of the total velocity dispersion can be written as

$$\sigma^2 = \frac{1}{2}\sigma_{12}^2(3 - \beta). \qquad (2)$$

Taking into account (2) and the fact that the pressure is $p \sim \rho\sigma^2$, it is clear that even a small increase in the anisotropy parameter $\beta$ gives some increase in the model's density $\rho$.

On a sample of 10 galaxies in (Kondratyev 1981), it was found that the anisotropy index weakly correlates with the average flattening of isophotes, and the upper anisotropy limit is $\beta_{max} \approx 0.38$. On a sample of 16 E-galaxies in (Caimmi 2009), these results were refined:

$$0.0 \leq \beta \leq 0.30, \tag{3}$$

where $\beta = 0$ corresponded to the galaxy NGC 4486 (E1), and $\beta = 0.30$ - to the galaxy NGC 4460 (E5.3).

The problem of velocity dispersion anisotropy was also discussed in (Tonry 1983), where it was noted that a radialy flattened velocity ellipsoid indirectly indicates to the origin of this E-galaxy through the merging of two spiral galaxies.

Over the past 10–15 years, new methods of observation have appeared (large ground-based telescopes and integral field spectrometers), with the help of which a lot of new information about elliptical galaxies was obtained. The quality of the published photometry of giant E-galaxies has improved markedly. In particular, isophote flattening was measured on the far periphery of giant E-galaxies, so the sizes of the studied regions in galaxies increased tenfold. The abundance of new information about the structure of E-galaxies attracted attention to the construction of their dynamic models again.

However, at this stage of research, new information on elliptical galaxies remains unclaimed to the full. In the project $ATLAS^{3D}$ (Cappellari 2008; Cappellari et al. 2013), Jeans Anisotropic Models (JAM models) were built, which were proved to be good at describing small galaxies with a regular velocity field and low anisotropy. However, this approach depends highly on kinematics information, and so only central parts of star systems may be studied by this way now. Closest giant ellipticals (NGC 4472, 4486) may have size 10 times larger than the area, for which good kinematical data is available. JAM models with rotational symmetry were also used to describe giant E-galaxies, which have a weak rotation and are likely triaxial or prolate.

Thus virial method, taking into account the data about the whole galaxy, may be useful for the estimation of the anisotropy of giant systems. But in the original version (Binney 1978) it uses only a model of an oblate spheroid with constant flattening throughout the system, ignoring the fact that ellipticity in real E-galaxies varies with radius significantly (see fig. 1).

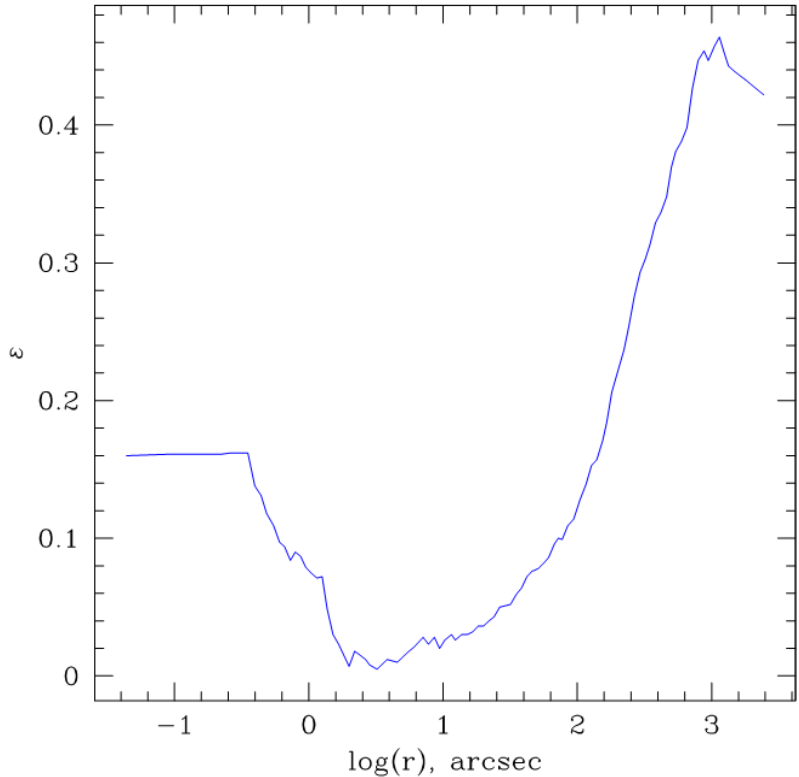

**Figure 1.** Profile of isophote flattening for the galaxy M87

In this regard, let us recall an important theoretical fact (Kondratyev 1982): *in inhomogeneous ellipsoids with similar density layers, the ratio of the rotational energy $T_{rot}$ to the module of the gravitational energy of the system $|W|$ (about the relation $t = \frac{T_{rot}}{|W|}$ see Section 2) does not depend on matter concentration at all and exactly coincides with the same ratio for homogeneous classical equilibrium figures.* Consequently, the models with such structure, tempting in their simplicity of use, however, do not allow finding the correct ratio values $t = \frac{T_{rot}}{|W|}$ for many galaxies. Thus, the original virial method should be modified to respond to the modern requirements for dynamical models of elliptical galaxies.

We emphasize that this work does not claim to be a complete review of methods for studying the dynamics of elliptical galaxies. It is focused mainly on the development and application of a method based on the virial tensor theorem. A key feature of these models is a rigorous consideration of the influence of the spatial shape of galaxies (oblate spheroid or triaxial ellipsoid) on the dynamics, as well as the structure of the internal density layers in these systems.

The outline of this paper is as follows. Section **2** describes our method. Sections **2.1-2.3** are concerned with the mathematical apparatus for calculating the relation $t = \frac{T_{rot}}{|W|}$. Sections **2.4-2.6** are concerned with the calculation of the anisotropy parameter of the velocity dispersion $\beta$ in galaxies. Section **3** presents a sample of elliptical galaxies selected here for the study. In Sections **3.1** and **3.2** the issues related to the spatial shape of galaxies and the dependence of the ratio $M/L$ with distance from the center are discussed. The results of modeling of the first 12 E-galaxies are presented in Section **4**. In Section **4.1**, we compare the results of calculation of the anisotropy parameter $\beta$ for models with a constant and variable profile of the isophotes. In Section **4.3**, our calculations are compared with the results obtained in the $ATLAS^{3D}$ project. Section **5** is devoted to comparing our results with the outcome of the high resolution cosmological simulations from the Illustris project. Section **6** summarizes the results of this work and discusses the correlation between anisotropy of giant galaxies and their outer ellipticity.

## 2. THE METHOD
### 2.1. Virial equations

Let us consider a layered-inhomogeneous ellipsoid, rotating with an angular velocity $\Omega$ around the axis $Ox_3$ (Kondratyev 1982, 1989, 2003). In this reference frame, the diagonal components of a second-order virial equation are

$$2T_{11} + W_{11} + \Pi_{11} + I_{11}\Omega^2 + 2\Omega \int_V \rho u_2 x_1 dV = 0,$$

$$2T_{22} + W_{22} + \Pi_{22} + I_{22}\Omega^2 - 2\Omega \int_V \rho u_1 x_2 dV = 0, \quad (4)$$

$$W_{33} + \Pi_{33} = 0.$$

Here

$$T_{ij} = \frac{1}{2}\int_V \rho u_i u_j dV; \quad \Pi_{ij} = \int_V \rho \left(\dot{x}_i - u_i\right)\left(\dot{x}_j - u_j\right) dV, \quad (5)$$

is a kinetic energy tensor of centroid motion and a tensor of energy of chaotic motion of the stars, respectively; $\boldsymbol{u}$ is the velocity vector of the ordered, and $(\dot{x} - \boldsymbol{u})$ - of the residual motions of the stars. There is no meridional circulation in the system, and thus $T_{33} = 0$.

The expression $2\Omega \int_V \rho u_i x_j dV$ describes the action of the Coriolis forces.

Equations (4) allow us to study the dynamics of stellar systems with both isotropic $\left(\Pi_{11} = \Pi_{22} = \Pi_{33}\right)$ and anisotropic $\left(\Pi_{11} \neq \Pi_{22} \neq \Pi_{33}\right)$ pressure. At this stage of the study, it is convenient to take the ratio

$$k = \frac{\Pi_{11} + \Pi_{22}}{2\Pi_{33}}. \quad (6)$$

as a measure of velocity dispersion anisotropy in the system.

A thorough analysis of the system of equations (4) has been given in the papers, mentioned above, and here we give only the formula for rotational energy of the configuration

$$T_{rot\,ob} = \frac{1}{2}\left(2kW_{33} - W_{11} - W_{22}\right). \tag{7}$$

In particular, $k = 1$, in a system with isotropic pressure, and then

$$T_{rot\,is} = \frac{1}{2}\left(2W_{33} - W_{11} - W_{22}\right). \tag{8}$$

Remark. The formula (7) for the rotation energy of the configuration is derived from the addition of the first two viral equations in (4) and the exclusion of $\Pi_{33}$ component using the third equation. Therefore, the kind of formula (7) does not depend on whether the configuration has only solid state rotation or only internal vortex movements, or both movements at the same time.

Note that the parameter $k$ from (6) can also be expressed through the ratio of the velocity dispersion in the equatorial plane $\sigma_{12}$ to the dispersion $\sigma_{33}$ along the rotational axis

$$k = \frac{\frac{1}{2}M\sigma_{12}^2}{M\sigma_{33}^2} = \frac{\sigma_{12}^2}{2\sigma_{33}^2}, \tag{9}$$

In its turn, the anisotropy parameter $\beta = 1 - \frac{\sigma_{33}^2}{\sigma_{12}^2}$ from (1) can be expressed in terms of $k$ (see below formula (31)).

**2.2. Gravitational energy tensor and total rotational energy of a layered inhomogeneous ellipsoid**

Following the works (Kondratyev 1982, 1989, 2003), we consider a layered inhomogeneous ellipsoid with the density distribution

$$\rho = \rho(m), \tag{10}$$

where

$$\frac{x_1^2}{a_1^2} + \frac{x_2^2}{\alpha_2^2(m)a_2^2} + \frac{x_3^2}{\alpha_3^2(m)a_3^2} = m^2, \quad 0 \leq m \leq 1. \tag{11}$$

Here, $m$ is the distribution parameter, $a_1, a_2, a_3$ denote semi-axes of the boundary ellipsoid. Level density layers are an ellipsoids with semi-axes $ma_1, m\alpha_2(m)a_2, m\alpha_3(m)a_3$. The functions $\alpha_i(m)$ define the type of stratification of the layers and must satisfy the following conditions:
1) they are continuous together with their first derivatives;
2) surfaces of equal density (11) should not intersect with each other;
3) on the surface of the boundary ellipsoid $\alpha_2(1) = \alpha_3(1) = 1$.

In the particular case of an ellipsoid with similar density layers, all $\alpha_i(m) = 1$. The mass and moments of inertia of the ellipsoidal subsystem with the surface $S(m)$ are

$$M(m) = \frac{4}{3}\pi a_1 a_2 a_3 \int_0^m dm\, \rho(m) \frac{d}{dm}\left[m^3 \prod_{i=1}^{} \alpha_i(m)\right];$$

$$I_{ij}(m) = \frac{4}{15}\pi a_1 a_2 a_3 a_i a_j \delta_{ij} \int_0^m dm\, \rho(m) \frac{d}{dm}\left[m^5 \alpha_i^2(m) \prod_{i=1}^{} \alpha_i(m)\right].$$

(12)

Further, the differences of the inertia moments will also be needed

$$N_{kl}(m) = I_{kk}(m) - I_{ll}(m);\quad k,l = 1,2,3.$$

(13)

The internal and external potentials of such layered inhomogeneous ellipsoids have been carefully studied in the works, mentioned above. A gravitational energy tensor $w_{ij}$ was also found there; through components $w_{ij}$ one can express the rotational energy in the form

$$T_{rot\ is} = T_1 - T_2 - T_3,$$

(14)

where

$$T_1 = \frac{1}{2}\pi G \int_0^1 dm\, \rho(m) M(m) \frac{d}{dm}\left[m^2\left(\alpha_1^2(m) a_1^2 A_1 + \alpha_2^2(m) a_2^2 A_2 - 2\alpha_3^2(m) a_3^2 A_3\right)\right];$$

$$T_2 = \frac{3}{2}\pi G \int_0^1 dm\, \rho(m) \left\{N_{13}(m) \frac{d}{dm}\left[\frac{A_1 - (1-e_{13}^2) A_3}{e_{13}^2} + 1\right] + N_{23}(m) \frac{d}{dm}\left[\frac{A_2 - (1-e_{23}^2) A_3}{e_{23}^2} + 1\right]\right\};$$

(15)

$$T_3 = \pi G \int_0^1 dm\, \rho(m) \left\{N_{21}(m) \frac{dA_1}{dm} + (I_{22}(m) + 2I_{33}(m)) \frac{dA_3}{dm}\right\}.$$

So total gravitational energy of the ellipsoid is equal to

$$W = -\pi G \int_0^1 dm\, \rho(m)\left[M(m) \frac{dI(m)}{dm} - N_{21}(m) \frac{dA_2}{dm} + N_{13}(m) \frac{dA_3}{dm}\right],$$

(16)

where

$$I(m) = m^2 \sum_{i=1}^{3} \alpha_i^2(m) a_i^2 A_i;$$

$$A_i(m) = a_1 a_2 a_3 \prod_{i=1}^{} \alpha_i(m) \int_0^\infty \frac{m^2 du}{\Delta(m,u)\left[m^2\left(\alpha_i^2(m) a_i^2 + u\right)\right]};$$

(17)

$$\Delta^2(m,u) = m^2 \prod_{i=1}^{}\left(\alpha_i^2(m) a_i^2 + u\right),$$

and $N_{kl}(m)$ is given in (13).

If the flattening of elementary layers of the equal density changes from the inner layers to the outer, then all the characteristics of the models: mass, gravitational energy, rotational energy, the ratio $t = \frac{T_{rot}}{|W|}$ and the ratio $\frac{v_{rot}}{\sigma}$ depend on the specific density profiles for this model. It is im-portant to note that the larger the concentration of the substance in the center, the greater these characteristics of the models depend on the density profile.

## 2.3. Using photometric data to calculate the ratio $t = \dfrac{T_{rot}}{|W|}$

Despite the high quality of modern photometric measurements, without measurements of the isophote flattening, these data are useless for studying the dynamics of galaxies by our method. In addition, due to the influence of the atmosphere, the blurring of the central part of the image cannot be avoided in ground-based observations. Therefore, only specific observational data are suitable for use in our models.

With this in mind, we chose the catalog (Kormendy et al. 2009). It contains surface brightness, flattening and position angle profiles for 43 early-type galaxies from Virgo cluster. The data were obtained by a combination of observations from ground-based telescopes and Hubble Space Telescope. Thus, they combine precision measurements in the central part of galaxies (with an angular resolution $< 0.03''$) and the data for the far edges of stellar systems. The ultimate magnitude from square second $\sim 26.5^m$.

The surface brightness profile of an elliptical galaxy is satisfactorily described by the well-known Hubble law (Hubble 1930), which is conveniently presented here as

$$I = \frac{I_0}{1 + \delta m^2}, \qquad (18)$$

where the parameter $m$ from (11) varies in the interval $0 \le m \le 1$. To make this law to account for the detailed structure of the surface brightness profile we added a varying coefficient $\delta$. If one writes (18) in the linear form

$$\frac{I_0}{I} = 1 + \delta m^2, \qquad (19)$$

it becomes clear that $\delta$ is the varying graph's slope.

To improve the accuracy of calculations in a layered-inhomogeneous model, the profile of each galaxy was divided into 8 - 15 sections, and $\delta$ was determined for each section. Depending on the size of the system, $\delta$ was chosen so that the likelihood coefficient for the linear form of the law (19) was $R^2 \sim 0.99$. A comparison of the observed profile with Vaucouleurs and Hubble laws with variable $\delta$ is shown in Fig. 2.

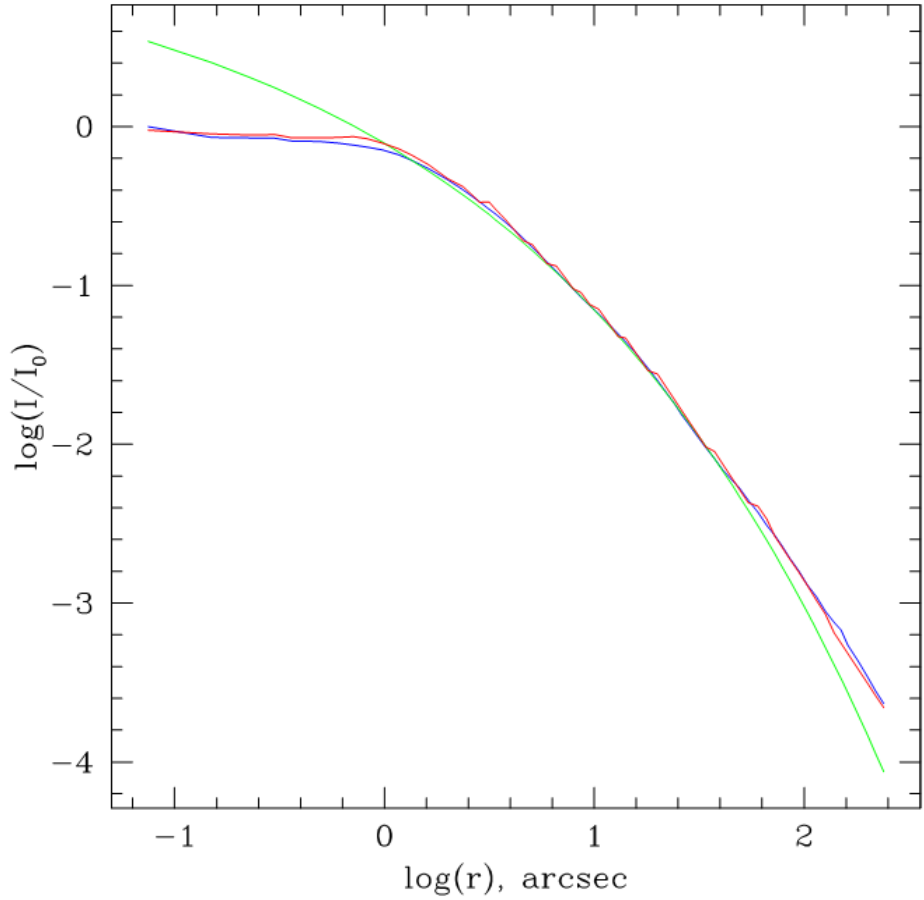

**Figure 2.** Comparison of the observed surface brightness profile for the galaxy NGC 4261 (blue line) with Vaucouleurs distribution (green line) and Hubble law with variable parameter $\delta$ (red)

As shown in Fig. 2, Hubble law with the variable coefficient $\delta$ is better than the Vaucouleurs law; it describes not only the central part of the galaxy, where most of the mass is concentrated, but also the margins, where the most significant changes in flattening of isophotes are observed.

To create a model, it suffices for us to know the relative density distribution. As it is known, the density profile $\rho(m)$ for the Hubble law (18) is

$$\rho(m) = \frac{\arctan\sqrt{\delta\frac{1-m^2}{1+\delta m^2}}}{(1+\delta m^2)^{\frac{3}{2}}}. \qquad (20)$$

**2.4. Velocity dispersion anisotropy parameter** $\beta = 1 - \dfrac{\sigma_{33}^2}{\sigma_{12}^2}$

Let us write the components of the energy tensor of chaotic motion of stars from (5) in the form

$$\begin{aligned}
\Pi_{11} &= \Pi + \tilde{\Pi}_{11}, \\
\Pi_{22} &= \Pi + \tilde{\Pi}_{22}, \\
\Pi_{33} &= \Pi + \tilde{\Pi}_{33}.
\end{aligned} \qquad (21)$$

Here $\Pi \approx \frac{1}{3} M \sigma^2$ is 2/3 of total internal energy of random motion of stars in the system. The anisotropy of pressure is taken into account by a tensor $\tilde{\Pi}_{ij}$, the trace (deviator) of which is zero.

Introducing now the notation

$$t_{is} = \frac{T_{rot\,is}}{|W|}; \quad t_{ob} = \frac{T_{rot\,ob}}{|W|}; \quad Q_3 = -\frac{\tilde{\Pi}_{33}}{T_{ob}}, \tag{22}$$

we obtain an important relationship from the equations (4)

$$\frac{3}{2} Q_3 = \frac{t_{is}}{t_{ob}} - 1, \tag{23}$$

Taking into account the known to us $T_{rot\,is}$ from (8) and $T_{rot\,ob}$ from (7), we write (23) in the form

$$\frac{3}{2} Q_3 = \frac{t_{is}}{t_{ob}} - 1 = \frac{2W_{33} - W_{11} - W_{22}}{2kW_{33} - W_{11} - W_{22}} - 1. \tag{24}$$

From (24) we find

$$2k\left(1 + \frac{3}{2} Q_3\right) W_{33} = \frac{3}{2} Q_3 \left(W_{11} + W_{22}\right) + 2W_{33}, \tag{25}$$

or, taking into account equations (5) again,

$$2k\left(1 + \frac{3}{2} Q_3\right) \Pi_{33} = \frac{3}{2} Q_3 \left(2T_{ob} + \Pi_{11} + \Pi_{22}\right) + 2\Pi_{33}. \tag{26}$$

From (26) follows

$$2k\left(1 + \frac{3}{2} Q_3\right) = 2 + \frac{3}{2} Q_3 \left(\frac{2T_{ob}}{\Pi_{33}} + 2k\right), \tag{27}$$

whence

$$2k = 2 + \frac{3}{2} Q_3 \frac{M v_{rot}^2}{\Pi + \tilde{\Pi}_{33}} = 2 + \frac{3}{2} Q_3 \frac{M v_{rot}^2}{\Pi \left(1 + \frac{\tilde{\Pi}_{33}}{\Pi}\right)}. \tag{28}$$

But since

$$\frac{\tilde{\Pi}_{33}}{\Pi} = \frac{\tilde{\Pi}_{33}}{T_{ob}} \frac{T_{ob}}{\Pi} = -\frac{3}{2} Q_3 \left(\frac{v_{rot}}{\sigma}\right)^2, \tag{29}$$

then (28) is reduced to

$$2k = 2 + \frac{\frac{9}{2} Q_3 \left(\frac{v_{rot}}{\sigma}\right)^2}{\left(1 + \frac{3}{2} Q_3 \left(\frac{v_{rot}}{\sigma}\right)^2\right)}. \tag{30}$$

Thus, according to formulas (9) and (30), the anisotropy parameter of the velocity dispersion (1) in the galaxy model can be written as

$$\beta = 1 - \frac{\sigma_{33}^2}{\sigma_{12}^2} = 1 - \frac{1}{2k} = \frac{1+4\kappa}{2+5\kappa}, \tag{31}$$

where we denoted

$$\kappa = \frac{3}{2} Q_3 \left( \frac{\upsilon_{rot}}{\sigma} \right)^2. \tag{32}$$

## 2.5. Scheme of calculating the anisotropy of velocity dispersion $\beta$

Next, our calculation plan is the following. Knowing $\upsilon_{rot}$ and $\sigma$ from observations allows, from the scalar virial theorem

$$|W| = M \cdot (\gamma \upsilon_{rot})^2 + M \sigma^2, \tag{33}$$

to find another necessary relation

$$t_{ob} = \frac{T_{rot\,ob}}{|W|} = \frac{1}{2} \frac{\left( \gamma \frac{\upsilon_{rot}}{\sigma} \right)^2}{1 + \left( \gamma \frac{\upsilon_{rot}}{\sigma} \right)^2}. \tag{34}$$

Note that the relation (34) describes each galaxy. Here $\gamma < 1$, due to the fact that $\upsilon_{rot}$ is a *maximum* rotational velocity. The average value $\gamma$ found from several rotation curves is approximately 0.87.

On the other hand, as shown in Section **2.3**, knowing the isophote flattening profile $\varepsilon(m)$, and the density distribution $\rho = \rho(m)$ from observations, taking into account the change in the mass-luminosity relation $M/L$ inside the galaxy (see Sec. **3.3**), we can calculate the value $t_{is}$ according to the formula

$$t_{is} = \frac{T_{rot\,is}}{|W|} = \frac{T_1 - T_2 - T_3}{|W|}, \tag{35}$$

Then, using formula (24), we find the very value of the parameter $Q_3$. Knowing $Q_3$ and the ratio $\left( \frac{\upsilon_{rot}}{\sigma} \right)_{ob}$, using formulas (31) and (32), we can now calculate the value of the velocity dispersion anisotropy parameter $\beta$ for each individual galaxy.

## 2.6. Influence of the orientation of the galaxy on the parameter $Q_3$

Let us suppose that an axisymmetric galaxy with an average flattening of isophotes $\varepsilon_{ob}$ is randomly oriented relative to the observer. As it is known from (Lindblad 1959), the observed and actual flattening of the galaxy $\varepsilon_{ob}$ and $\varepsilon_{true}$ are related to the position angle $i$ of its equatorial plane by the formula

$$\cos i = \left[ \frac{\varepsilon_{ob}(2-\varepsilon_{ob})}{\varepsilon_{true}(2-\varepsilon_{true})} \right]^{\frac{1}{2}}. \tag{36}$$

The ratio of the observed rotational velocity of the galaxy $\upsilon_{ob}$ to the true one $\upsilon_{true}$ is also equal to $\cos i$. As $\varepsilon_{ob} \leq \varepsilon_{true}$, then $\left(\frac{\upsilon_{rot}}{\sigma}\right)_{ob} \leq \left(\frac{\upsilon_{rot}}{\sigma}\right)_{true}$. Therefore, according to the formula (34), the inequality $t_{true} \geq t_{ob}$ will be fulfilled. The last inequality gives an important restriction (see formula (23)) to $Q_3$:

$$(Q_3)_{ob} \leq (Q_3)_{true}. \tag{37}$$

This inequality, in its turn (see formula (32)), means that $\kappa_{ob} \leq \kappa_{true}$. Now, by virtue of relation (31), we conclude

$$\beta_{true} \geq \beta_{ob}. \tag{38}$$

However, it must be emphasized that the orientation of specific triaxial E- galaxies relative to the observer (three Euler angles) is not known to us in detail. Therefore, here we do not pretend to resolve completely the question of the influence of the orientation of E- galaxies on the anisotropy of velocity dispersion. Inequality (38) gives only an estimate for the lower limit of the velocity dispersion anisotropy, and only for axisymmetric elliptical galaxies.

## 3. APPLICATION OF OBSERVATIONAL DATA

### 3.1. Galaxies under study

Since we are mostly interested in those elliptical galaxies, in whose dynamics the anisotropy of velocity dispersion plays a significant role, all E-galaxies with stellar mass $M_* \geq 10^{11} M_{Sun}$ were selected from the catalog, as well as other available E-galaxies, which were classified as slow rotators in (Emsellem et al. 2011). Two rapidly rotating galaxies, NGC 4434 and NGC 4621, whose velocity distribution is expected be close to isotropic, were also included in the sample to test the method. A small E-galaxy NGC 4473 was added to this list, because it shows rapid rotation, but its magnitude is not enough to maintain its highly flattened shape.

The main properties of the 12 studied galaxies are given in Table 1.

| Name | $R_e$, kpc | $V_{rot}/\sigma$ | $\varepsilon_e$ | $\log M_*$, $M_{Sun}$ | F/S |
|---|---|---|---|---|---|
| NGC 4261 | 5.85 | 0.081 | 0.222 | 11.42 | S |
| NGC 4365 | 5.59 | 0.100 | 0.254 | 11.37 | S |
| NGC 4374 | 4.33 | 0.024 | 0.147 | 11.35 | S |
| NGC 4472 | 7.49 | 0.063 | 0.172 | 11.69 | S |
| NGC 4486 | 6.78 | 0.021 | 0.037 | 11.5 | S |
| NGC 4552 | 2.57 | 0.047 | 0.047 | 11.05 | S |
| NGC 4636 | 6.98 | 0.036 | 0.094 | 11.09 | S |
| NGC 4649 | 5.33 | 0.107 | 0.156 | 11.41 | F |
| NGC 4458 | 2 | 0.143 | 0.121 | 10.04 | S |
| NGC 4621 | 3.24 | 0.253 | 0.365 | 10.97 | F |
| NGC 4434 | 1.55 | 0.227 | 0.058 | 10.37 | F |
| NGC 4473 | 2.13 | 0.252 | 0.421 |  | F |

**Table 1.** Main characteristics of the studied galaxies: (2) $R_e$ is the effective radius; (3) is the relation $V_{rot}/\sigma$ on $R_e$; (4) $\varepsilon_e$ is the flattening on $R_e$; (5) $\log M_*$ is the logarithm of the total stellar mass in the masses of the Sun; (6) - classification of the galaxy according to the $ATLAS^{3D}$ scheme (F - fast, S - slow rotator); data of columns (2, 3, 5) are taken from the catalog (Dabringhausen & Fellhauer 2016), (4, 6) - from (Emsellem et al. 2011)

### 3.2. The results of the observational tests

In the present work, the galaxies were simulated with either a rotationally symmetric oblate spheroid, or with a prolate spheroid without the symmetry around the axis of rotation $Ox_3$. There is not enough data to use a triaxial ellipsoid model, however, the dynamic properties of the prolate spheroid with relatively large flattening are close to those of the triaxial ellipsoid (Kondratyev 2003).

To determine which model should describe a concrete galaxy, the results of the observational tests (Kondratyev & Ozernoy 1979) were used. First of all, we used the data on the deviation of the projected rotational axis from the observed minor axis of the galaxy (Krajnović et al. 2011) and the position angle profiles of 12 systems under our study (see Tables 2-3).

| Name | Ψ, ° | Name | Ψ, ° | Name | Ψ, ° |
|---|---|---|---|---|---|
| NGC 4261 | 73.7 | NGC 4486 | 46.2 | NGC 4458 | 20.1 |
| NGC 4365 | 75.9 | NGC 4552 | 7.2 | NGC 4621 | 2.0 |
| NGC 4374 | 42.7 | NGC 4636 | 57.2 | NGC 4434 | 7.7 |
| NGC 4472 | 14.3 | NGC 4649 | 0.2 | NGC 4473 | 0.2 |

**Table 2.** Deviation of the projected rotational axis from the visible minor axis of the elliptical limb (also known as kinematic misalignment) for the concrete galaxy: here $\Psi = \varphi - \varphi'$ ($\varphi$ is the position angle of the observed minor axis and $\varphi'$ is the position angle of the projected axis of rotation), according to (Krajnović et al. 2011). The data cover only the central regions of galaxies.

| Name | Δφ,° | Name | Δφ,° | Name | Δφ,° |
|---|---|---|---|---|---|
| NGC 4261 | 48.7 | NGC 4486 | 197.7 | NGC 4458 | 6.8 |
| NGC 4365 | 11.3 | NGC 4552 | 41.8 | NGC 4621 | 3.5 |
| NGC 4374 | 69.8 | NGC 4636 | 23.7 | NGC 4434 | 8.4 |
| NGC 4472 | 37.3 | NGC 4649 | 42.7 | NGC 4473 | 4.0 |

**Table 3.** The difference between the maximum and minimum value of the position angle of the observed minor axis of the galaxy, from (Kormendy et al. 2009)

As it can be seen, the galaxies NGC 4261, 4365, 4374, 4486 and 4636 have a strong mismatch between the projection of the rotation axis and the observed minor axis of the limb. The galaxies NGC 4472 and 4552 show a slight misalignment between axes, but at the same time they are giant systems and demonstrate a significant disruption in isophote coaxiality. For all these seven galaxies, prolate spheroid model was used. Galaxy NGC 4649 shows a slight isophote twist and a slight mismatch of the axes, although it is a giant system, so we modeled it with both methods. The rest of the galaxies: NGC 4458, 4621, 4434 and 4473 do not show a large misalignment of axes or isophote twisting. An oblate spheroid model was used for them.

### 3.3. Gradient profile of mass-luminosity relation $M/L$

The initial mass distribution, with which stars are formed, is a fundamental property of the galaxy. It influences on everything that can be directly observed in a star system. The initial mass function (IMF) of the Milky Way has been studied quite well and is described by a simple law (Kroupa 2001; Cappellari et al. 2013), but the situation may be different in other galaxies. Independent methods, based on gravitational lensing and stellar kinematics (Auger 2010; Cappellari et al. 2012), confirm that massive E-galaxies have a greater stellar mass, than predicted by the IMF of the Milky Way. Van Dokkum & Conroy (2010) suggested that the difference in the IMF is associated with an excess of low-mass stars in giant E-galaxies. Recent detailed analysis (Masi et al. 2018) shows that observed distributions of metal abundances in stars of giant E-galaxies are better described by IMF, which changes its type during the evolution of the system.

It was previously considered (Strom & Strom 1978) that in elliptical galaxies the mass-to-luminosity ratio $M/L$ is constant throughout. However, now this assumption needs to be questioned in the light of the aforementioned studies and with the approval of a new paradigm about the formation of massive E-galaxies. It says that the central, dense part of such systems was formed in the early universe during in situ star formation episode, while the marginal part appeared later as a result of accretion of satellites, which continues to this day (Naab et al. 2009). Based on this assumption, analyzing kinematics of stars, globular star clusters and satellites of the galaxy M87, the authors Oldham & Auger (2018) concluded that its stellar ratio $M/L$ is described by the law

$$\Gamma^* \sim R^\mu, \quad \mu = -0.54 \pm 0.05, \tag{39}$$

where $R$ is the distance from the center of the system.

Considering the above, we have included the gradient of the mass-luminosity ratio in the model of M87, since it takes into account the additional change in bulk density over the galaxy body and indirect influence of the supermassive black hole in the center.

The introducing of the gradient $M/L$ profile appears on the stage of calculating the density profile (see Sec **2.3**), because the initial relation (20) was obtained, using $M/L = 1$ assumption. Note, that the new gradient profile accounts for the stellar matter only, we do not use any suppositions about dark matter halo.

## 4. RESULTS OF MODELING
### 4.1. Comparing models with variable and constant flattening of density layers

The results of calculations of the parameter $3/2Q_3$ from (23) and its corresponding anisotropy index $\beta$ from (31) for a model with similar layers and a more realistic model with variable flattening $\varepsilon(m)$ are shown in Table 4. Two values of anisotropy parameters correspond to each galaxy: the first for $\gamma = 1$ and the second for $\gamma = 0.87$.

The uncertainties of computed anisotropy values have a statistical nature and originate from observational errors in kinematics (taken from (Dabringhausen & Fellhauer 2016)) and photometry (from (Kormendy et al. 2009) and MIDAS software (Bender & Moellenhoff 1987)). Systematic errors, although present, are difficult to estimate (more in Section 5).

| Name | $\varepsilon_e$ | $\varepsilon$ | $3/2Q_3$ similar | $\beta$ similar | $\Delta\beta$ similar | $3/2Q_3$ | $\beta$ | $\Delta\beta$ |
|---|---|---|---|---|---|---|---|---|
| NGC 4261 prol. | 0.222 | 0.152 | 9.2 | 0.080 | 0.009 | 10.6 | 0.090 | 0.007 |
| | | 0.376 | 12.4 | 0.103 | | 14.3 | 0.116 | |
| NGC 4365 prol. | 0.254 | 0.03 | 7.0 | 0.089 | 0.012 | 11.7 | 0.135 | 0.032 |
| | | 0.376 | 9.6 | 0.115 | | 15.8 | 0.169 | |
| NGC 4374 prol. | 0.147 | 0.024 | 75.5 | 0.057 | 0.004 | 20.3 | 0.016 | 0.004 |
| | | 0.184 | 100 | 0.073 | | 27.1 | 0.022 | |
| NGC 4472 prol. | 0.172 | 0.063 | 11.7 | 0.063 | 0.037 | 19.0 | 0.096 | 0.035 |
| | | 0.27 | 15.7 | 0.082 | | 25.5 | 0.122 | |
| NGC 4486 prol. | 0.037 | 0.005 | 22.9 | 0.014 | 0.008 | 394 | 0.176 | 0.029 |
| | | 0.464 | 30.6 | 0.019 | | 522 | 0.213 | |
| NGC 4552 prol. | 0.047 | 0.037 | 5.0 | 0.016 | 0.007 | 31.1 | 0.087 | 0.011 |
| | | 0.294 | 6.9 | 0.022 | | 41.4 | 0.110 | |
| NGC 4636 prol. | 0.094 | 0.015 | 19.6 | 0.014 | 0.007 | 99.3 | 0.145 | 0.020 |
| | | 0.445 | 26.3 | 0.047 | | 132 | 0.178 | |
| NGC 4649 obl. | 0.156 | 0.035 | 6.9 | 0.099 | 0.010 | 8.1 | 0.114 | 0.006 |
| | | 0.215 | 9.4 | 0.127 | | 11.0 | 0.144 | |
| NGC 4649 prol. | 0.156 | 0.035 | 3.1 | 0.049 | 0.010 | 3.8 | 0.060 | 0.008 |
| | | 0.215 | 4.2 | 0.065 | | 5.4 | 0.080 | |
| NGC 4458 obl. | 0.121 | 0.056 | 2.4 | 0.049 | 0.008 | 0.9 | 0.027 | 0.005 |
| | | 0.588 | 3.5 | 0.092 | | 1.5 | 0.044 | |
| NGC 4621 obl. | 0.365 | 0.114 | 2.9 | 0.185 | 0.013 | 0.12 | 0.012 | 0.018 |
| | | 0.366 | 4.1 | 0.237 | | 0.46 | 0.041 | |
| NGC 4434 obl. | 0.058 | 0.05 | | | 0.009 | | | 0.016 |
| | | 0.433 | -0.2 | -0.016 | | 0 | 0.000 | |
| NGC 4473 obl. | 0.421 | 0.289 | 3.7 | 0.234 | 0.012 | 2.2 | 0.156 | 0.019 |
| | | 0.688 | 5.1 | 0.268 | | 3.2 | 0.202 | |

**Table 4.** Results of calculation of anisotropy indices: (4, 5) - using the model with similar layers; (7.8) - using the model with variable flattening $\varepsilon(m)$; (6, 9) - uncertainties of computed anisotropy values originating from observational errors; obl. - oblate spheroid, prol. - prolate spheroid; (2) - flattening on $R_e$ from Emsellem et al. (2011); (3) — the minimum (above) and maximum (below) flattening of the galaxy according to data from Kormendy et al. (2009). Two values of anisotropy indices correspond to each galaxy: the first for $\gamma = 1$ and the second for $\gamma = 0.87$

On a sample of 12 E-galaxies, we found that the anisotropy parameter lies in the range

$$0.0 \leq \beta \leq 0.213, \tag{40}$$

Here $\beta = 0.0$ corresponds to a regular fast rotator NGC 4434 and $\beta_{max} = 0.213$ - to the galaxy M87, whose flattening increases rapidly from the center to the periphery. Comparing columns 5 and 8 in the Table 4, we see that accounting for the isophote flattening profile significantly affects the magnitudes of velocity dispersion anisotropy in many galaxies.

Analyzing the results in Table 4, it can be noted that the anisotropy parameters, obtained, using models with variable flattening and without it, differ significantly in cases when the flattening of the galaxy varies over a wide range. For example, the galaxies NGC 4486, 4636 and 4552 have a significant increase in flattening outside the central part. A model with similar layers, which ignores the real isophote profile, gives underestimated anisotropy parameter $\beta$ for these galaxies. At the same time, observational tests and data on the mass of the systems show that all these galaxies are giant triaxial systems, and therefore it is logical to expect them having high anisotropy of velocity dispersion. And it is important to emphasize that the model with variable flattening, which most fully takes into account the detailed structure of galaxies, gives exactly this result.

Note that two other examples also confirm the adequacy of the method, used here. The galaxy NGC 4621, on the contrary, has a decrease in flattening at distances more, than $1R_e$. This is a fast rotator, and observational tests suggest that its shape should be close to an oblate spheroid, so high anisotropy rates are not expected for this galaxy. Our model with variable flattening shows exactly such low values of anisotropy, in contrast to the model with similar layers. The other galaxy NGC 4458 is classified as a slow rotator, but it is not a giant system and has a regu-lar velocity field. It shows a slight disturbance in the isophote coaxiality, which may mean only a small deviation from the shape of the oblate spheroid. A model with similar layers gives too high anisotropy values for this galaxy, while the one, taking into account variable flattening, allows us to obtain a model that is close to isotropic.

Interestingly, the model with constant flattening of the layers gives a negative anisotropy index for the NGC 4434 galaxy. That is because the abnormally low oblateness at the effective radius does not reflect the dynamics of this rapidly rotating system as a whole.

A special case is the giant galaxy NGC 4374, for which relatively high anisotropy values can be expected, however, it has extremely weak rotation and has very little flattening beyond the effective radius. Either this galaxy is really close to isotropic, or it is lenticular, observed facet on, as reported by some authors (Emsellem et al. 2011; Nilson 1973).

Our method describes well the dynamics of those galaxies that do not have a noticeable skew in the orientation of the axis of rotation of the core. In this regard, we note that in our sample there are only two galaxies (NGC 4365 and NGC 4472) that have a noticeable rotation in the inclination of the axis of rotation of the nucleus.

An interesting example is the galaxy NGC 4473, for which the both models (axisymmetric with similar layers and with variable flattening) show strong anisotropy. It is a medium sized galaxy with signs of a fast rotator; however, isotropic system with such an oblateness should have $V_{rot}/\sigma \sim 0.8$, which means that it is impossible to explain the shape of this galaxy only by rotation. The authors in (Alabi et al. 2015) report that isophote twisting and anomalous deviation of the rotational axis is observed in the outer layers of this galaxy (beyond the sensitivity limits of spectrometers from the $ATLAS^{3D}$ project. Measurements of the velocity dispersion show that, instead of one peak in the galaxy, there are two areas with increased dispersion at a large distance from the center. Thus, this system belongs to a rare type of 2-sigma galaxy or a galaxy with a kinematically distinguished halo. It is also interesting to note that dynamics of the galaxy NGC 4473 was also modeled with 2 counter-rotating disk-like structures (see, for example, Cappellari & McDermid 2005; Cappellari et al. 2007; Foster et al. 2013).

## 4.2. $\frac{M}{L}$ gradient model

For the galaxy M87 we performed calculations, using not only the standard model with variable flattening, but the combined model as well, which also includes the gradient profile of the stellar mass-luminosity ratio. Taking into account the changes in the mass-luminosity ratio $M/L$ over the M87 body gives a decrease of the anisotropy by $21\%$.

It is characteristic that the addition of gradient $M/L$ profile gives a systematic decrease in the anisotropy parameter in the galaxy. As noted in the Introduction, this is due to the actual increase of the concentration of stellar matter in the central part of the system. By the way, such an effect can be obtained only in the model with variable flattening of the layers, since the influence of the density profile disappears in the models with similar layers.

### 4.3. Comparison with results of $ATLAS^{3D}$ project

In this part of the work we present the results of a comparison of the anisotropy parameters, obtained by our method, with the values, given in the project $ATLAS^{3D}$ (see Table 5)

| Name | $\varepsilon_e$ | $\varepsilon$ | $\beta$ | $\beta$ similar | $ATLAS^{3D}$ |
|---|---|---|---|---|---|
| NGC 4261 prol. | 0.222 | 0.152 | 0.086 | 0.080 | 0.02 |
|  |  | 0.376 | 0.110 | 0.103 |  |
| NGC 4365 prol. | 0.254 | 0.03 | 0.135 | 0.089 | 0.10 |
|  |  | 0.376 | 0.169 | 0.115 |  |
| NGC 4374 prol. | 0.147 | 0.024 | 0.016 | 0.057 | 0.05 |
|  |  | 0.184 | 0.022 | 0.073 |  |
| NGC 4472 prol. | 0.172 | 0.063 | 0.096 | 0.063 | 0.14 |
|  |  | 0.27 | 0.122 | 0.082 |  |
| NGC 4486 prol. | 0.037 | 0.005 | 0.176 | 0.014 | 0.00 |
|  |  | 0.464 | 0.213 | 0.019 |  |
| NGC 4552 prol. | 0.047 | 0.037 | 0.087 | 0.016 | 0.02 |
|  |  | 0.294 | 0.110 | 0.022 |  |
| NGC 4636 prol. | 0.094 | 0.015 | 0.145 | 0.014 | 0.07 |
|  |  | 0.445 | 0.178 | 0.047 |  |
| NGC 4649 obl. | 0.156 | 0.035 | 0.114 | 0.099 | 0.02 |
|  |  | 0.215 | 0.144 | 0.127 |  |
| NGC 4649 prol. | 0.156 | 0.035 | 0.060 | 0.049 |  |
|  |  | 0.215 | 0.080 | 0.065 |  |
| NGC 4458 obl. | 0.121 | 0.056 | 0.027 | 0.049 | 0.04 |
|  |  | 0.588 | 0.044 | 0.092 |  |
| NGC 4621 obl. | 0.365 | 0.114 | 0.012 | 0.185 | 0.05 |
|  |  | 0.366 | 0.041 | 0.237 |  |
| NGC 4434 obl. | 0.058 | 0.05 |  |  | 0.00 |
|  |  | 0.433 | 0.000 | -0.016 |  |
| NGC 4473 obl. | 0.421 | 0.289 | 0.156 | 0.234 | 0.00 |
|  |  | 0.688 | 0.202 | 0.268 |  |

**Table 5.** Comparison of anisotropy parameters, obtained in this work, with the results of the $ATLAS^{3D}$ project: (2) - flattening of the galaxy at the effective radius, according to data from Emsellem et al. (2011); (3) - the minimum and maximum flattening of the galaxy, according to data from Kormendy et al. (2009); (4) is the anisotropy parameter for the model with variable flattening; (5) - according to the model with constant oblateness; (6) is the anisotropy parameter obtained in $ATLAS^{3D}$ project (Cappellari 2016).

First of all, we note that, although the $ATLAS^{3D}$ results (Cappellari 2016) do not radically deviate from our results, however, a comparison with our models reveals a number of significant differences in the anisotropy parameters of the velocity dispersion. The reason for this is understandable, since only our models take into account galaxy's isophote structure from the center to the edges, where intracluster light starts to dominate. It should not be forgotten that we use not only rotationally symmetric models. That is why there is a noticeable discrepancy in the anisotropy parameters for the NGC 4486, 4552 and 4636 galaxies, similar to the discrepancy with the result of calculations, using a model with similar layers. These three star systems have very low flattening in the central areas and $ATLAS^{3D}$ team used exactly these values for their JAM models. But actually the flattening of these galaxies increases fast beyond $1R_e$, and, in addition, their spatial shape is not likely described by an oblate spheroid due to very weak rotation. The opposite is true for NGC 4374, the flattening of which decreases from center to the edge. That is why $ATLAS^{3D}$ gives higher magnitude of velocity dispersion anisotropy for this galaxy than our model.

As can be seen from the Table 6, for small galaxies with a regular velocity field, such as NGC 4458, 4434 and 4621, the results of Jeans anisotropic models is similar to ours. Such a comparison indicates the adequacy of our approach, since JAM models were developed exactly for such type of galaxies.

For the galaxies NGC 4472, 4365, 4649, we used the model of a prolate spheroid. Relatively small differences in anisotropy indices for these galaxies are due to the fact that high values of flattening at the effective radius reflect well the shape of the isophotes in the outer parts of these galaxies.

A prolate spheroid model has also been applied to the galaxy NGC 4261. We did this choice of model in accordance with the observational tests (Kondratyev & Ozernoy 1979). But there is a contradiction in the $ATLAS^{3D}$ approach: although it has been noted that this galaxy does not have rotational symmetry (Emsellem et al. 2011), however, the same model with axial symmetry was used in calculations in the (Cappellari 2016). This explains the strong difference with our results for this galaxy.

Note that, although $ATLAS^{3D}$ group states in (Emsellem et al. 2011) that their method may be inaccurate for some galaxies, they only pay attention to the galaxies with prolate rotation and lenticular galaxies that are visible face on or likely have bars. In the file, provided on their website with the anisotropy values, obtained using JAM models (which were published in (Cappellari 2016)), they warn that these values cannot be used, if the angle of inclination of the galaxy is very different from 90 degrees (i.e. if the galaxies is likely visible face on) and if the quality of the model is indicated as low (0 on the four-point scale). For all the galaxies from our sample (except for NGC 4636), the quality of the models is indicated as good (2) or satisfactory (1). We did not find any warnings about the use of anisotropy indicators for massive galaxies, as well as no mention of the problems of their method due to the small size of the study area and the use of an oblate spheroid model. By the way, the groups from MASSIVE (Veale et al. 2016) and SLUGGS surveys (Brodie et al. 2014) also criticize $ATLAS^{3D}$ for the overly limited area of kinematic analysis, which barely reaches the size of $1R_e$ in the case of massive galaxies. Thus, the results of $ATLAS^{3D}$ cannot be taken as average anisotropy over the entire galaxy in the case of massive systems.

### 4.3. Additional galaxies and the results of statistical study

The initial sample of galaxies should be increased to perform at least a simple statistical analysis. We decided to include another 10 giant galaxies from MASSIVE survey (Ma et al. 2014) in our study. We used the model of a prolate spheroid for all of them since they are extremely massive. Only the galaxies NGC 315 and NGC 1700, demonstrating relatively strong rotation and no signs of triaxiality, were also modeled by an oblate spheroid.

Photometric properties of these galaxies were obtained solely on Hubble telescope (Goullaud et al. 2018), since they are remote systems and have small angular sizes. Observational errors are slightly different from galaxies from Kormendy's catalog: kinematical errors are smaller due to using of modern VIRUS-P spectrograph (Murphy et al. 2011), while photometric measurments has a bit worse quality because of different procedure, chosen for ellipse fitting (Jedrzejewski, 1987)

The properties of additional galaxies are listed in Table 6, along with computed values of velocity dispersion anisotropy and their uncertanties.

| Name | $\varepsilon_e$ | $\varepsilon$ | $V_{rot}/\sigma$ | $\log M_*$, $M_{Sun}$ | $3/2 Q_3$ | $\beta$ | $\Delta\beta$ |
|---|---|---|---|---|---|---|---|
| NGC 315 obl. | 0.262 | 0.154 | 0.137 | 12.03 | 7.4 | 0.154 | 0.030 |
| | | 0.307 | | | 10.0 | 0.192 | |
| NGC 315 prol. | 0.262 | 0.154 | 0.137 | 12.03 | 3.6 | 0.088 | 0.021 |
| | | 0.307 | | | 5.1 | 0.116 | |
| NGC 410 prol. | 0.314 | 0.00 | 0.085 | 11.86 | 12.4 | 0.110 | 0.026 |
| | | 0.323 | | | 16.7 | 0.139 | |
| NGC 533 prol. | 0.260 | 0.104 | 0.042 | 11.92 | 64.3 | 0.134 | 0.023 |
| | | 0.339 | | | 85.3 | 0.166 | |
| NGC 777 prol. | 0.180 | 0.062 | 0.101 | 11.87 | 4.7 | 0.064 | 0.024 |
| | | 0.184 | | | 6.5 | 0.085 | |
| NGC 1016 prol. | 0.041 | 0.041 | 0.030 | 12.05 | 13.3 | 0.017 | 0.012 |
| | | 0.125 | | | 17.9 | 0.023 | |
| NGC 1060 prol. | 0.233 | 0.055 | 0.021 | 11.90 | 141 | 0.079 | 0.016 |
| | | 0.279 | | | 187 | 0.100 | |
| NGC 1272 prol. | 0.038 | 0.01 | 0.020 | 11.81 | 7.4 | 0.004 | 0.002 |
| | | 0.197 | | | 10.1 | 0.006 | |
| NGC 1600 prol. | 0.294 | 0.232 | 0.029 | 11.90 | 119 | 0.122 | 0.036 |
| | | 0.384 | | | 157 | 0.166 | |
| NGC 1700 | 0.296 | 0.192 | 0.202 | 11.72 | 3.5 | 0.157 | 0.056 |
| obl. | | 0.339 | | | 4.9 | 0.200 | |
| NGC 1700 prol. | 0.296 | 0.192 | 0.202 | 11.72 | 1.4 | 0.077 | 0.024 |
| | | 0.339 | | | 2.2 | 0.110 | |
| NGC 4914 prol. | 0.390 | 0.212 | 0.050 | 11.78 | 71.9 | 0.186 | 0.030 |
| | | 0.483 | | | 95.3 | 0.224 | |

**Table 6.** Main characteristics and results of calculation of velocity dispersion anisotropy for the galaxies from MASSIVE survey: (2) $\varepsilon_e$ is the flattening on $R_e$; (3) - the minimum and maximum flattening of the galaxy ; (4) is the relation $V_{rot}/\sigma$ on $R_e$; (5) $\log M_*$ is the logarithm of the total stellar mass in the masses of the Sun; (6) - parameter $3/2Q_3$ of anisotropy of velocity dispersion according to our model with variable flattening; (7) - corresponding parameter beta of anisotropy of velocity dispersion; (8) - statistical errors of parameter beta originating from observational errors; data of columns (2, 4) are taken from (Veale et al. 2016), (3) - from (Goullaud et al. 2018), (5) - from (Veale et al. 2018).

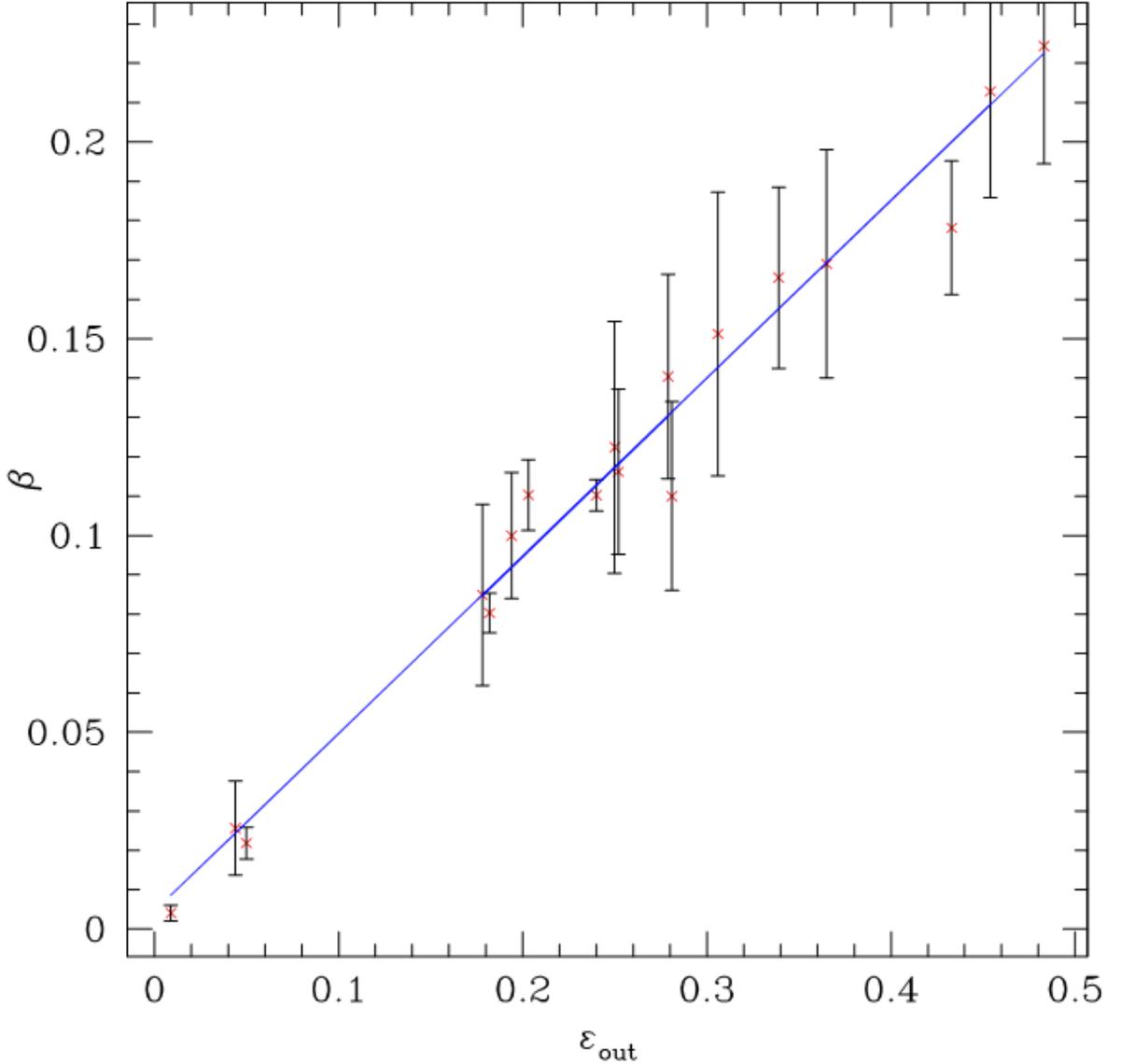

**Figure 3.** Dependence "anisotropy–flattening" for the outer regions of 18 giant elliptical galaxies. Anisotropy was calculated, using a prolate spheroid model with $\gamma$ – correction for the orientation of the galaxy. The $\varepsilon_{out}$ is taken as the average value of the flattening of the isophotes at a distance from $0.8R_{tot}$ to $R_{tot}$, where $R_{tot}$ is the total distance from the center to the edge. The error bars are given in the plot

Analyzing the widen sample, we found a correlation between the anisotropy of velocity dispersion and the flattening of the outer regions of E-galaxies (Fig. 5). The existence of this correlation confirms the conclusion that the anisotropy of velocity dispersion plays an important role in the dynamics of elliptical galaxies. In addition, the existence of the discussed correlation in the outer layers of elliptical galaxies may directly indicate that these parts of giant stellar systems were likely formed by an accretion of satellites.

## 5. COMPARISON WITH RESULTS of COSMOLOGICAL SIMULATIONS IN THE ILLUSTRIS PROJECT

Project Illustris (Vogelsberger et al. 2014b), we chose, was started in 2014 and focuses on the properties of the galaxies and clusters. Relatively small volume of the simulated universe (105 $Mpc^3$) provides an optimal amount of data and enough diversity of objects at the same time.

The anisotropy of velocity dispersion cannot be measured directly, since it is possible to obtain only line-of-sight velocities of stars. One of solutions is to turn to cosmological hydrodynamical simulations. Most of the galaxies, obtained in such projects, could be "observed" from multiple directions, and, consequently, we can get information about their velocity dispersion tensor. We extracted dynamical properties of the most massive ($\log M^* > 11$) early-type galaxies from the catalog (Xu et al., 2017) and, after a removal of lenticular galaxies, obtained a sample of 60 systems. Note, that, unfortunately, there is no data $z = 0$, and we have to use the data, corresponding to $z = 0.1$, the closest to the modern era. Although such difference should not lead to drastic discrepancy between results, we still cannot perform any fine analysis. Detailed data on the objects of this sample are collected in Table 7.

For the purposes of this article, we are interested in the line-of-sight velocity dispersions measured from the galaxy's images obtained in three different projections. Using this data for each massive galaxy we calculated the anisotropy parameter of the velocity dispersion, according to the formula (1).

The distribution of the sampled galaxies on the anisotropy parameter is shown in Fig. 3. The data is from (Xu D. et al. 2017). It is important that the anisotropy magnitudes turned out to be numerically quite close to our results. Indeed, we see that the parameters $\beta$ for the most of galaxies lie within the interval $[0 \div 0.3]$ and only for some of them this parameter turns out to be slightly larger. While anisotropy of the real galaxies computed with our method appeared in the interval $[0 \div 0.224]$ (considering that our sample is sufficiently smaller and we have not access to three projections of every galaxy). Thus, most of massive E-galaxies do have anisotropy, like real systems in our study.

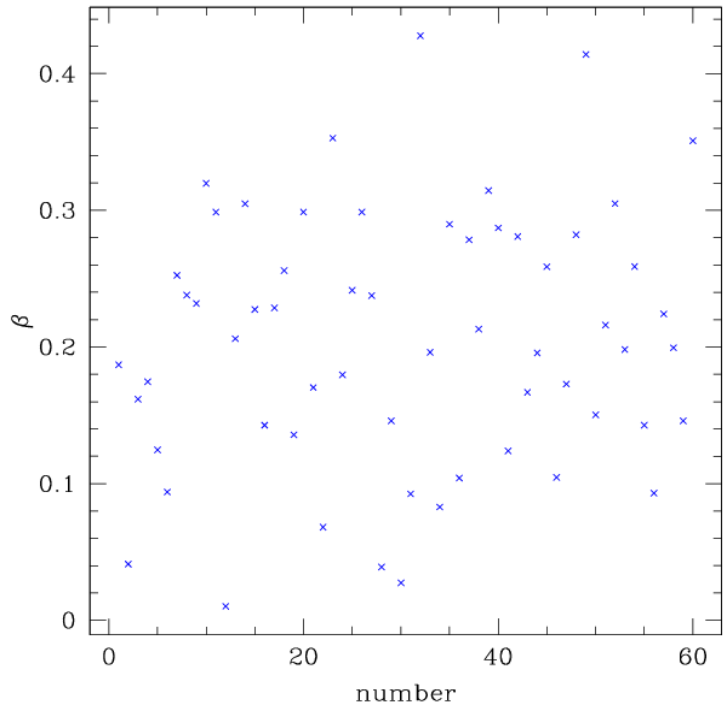

**Figure 4.** The distribution of the examined galaxies by the anisotropy parameter $\beta$. The x-axis is the number of the galaxy in the sample. The y axis is the anisotropy parameter $\beta$, calculated, using the formula (1). It is seen that most of the galaxies from the sample have anisotropy $\beta$ in the range of $0.1 \div 0.3$.

In addition, Fig.4 shows the dependence of the velocity dispersion anisotropy parameter on the mass of giant galaxies from the Illustris project.

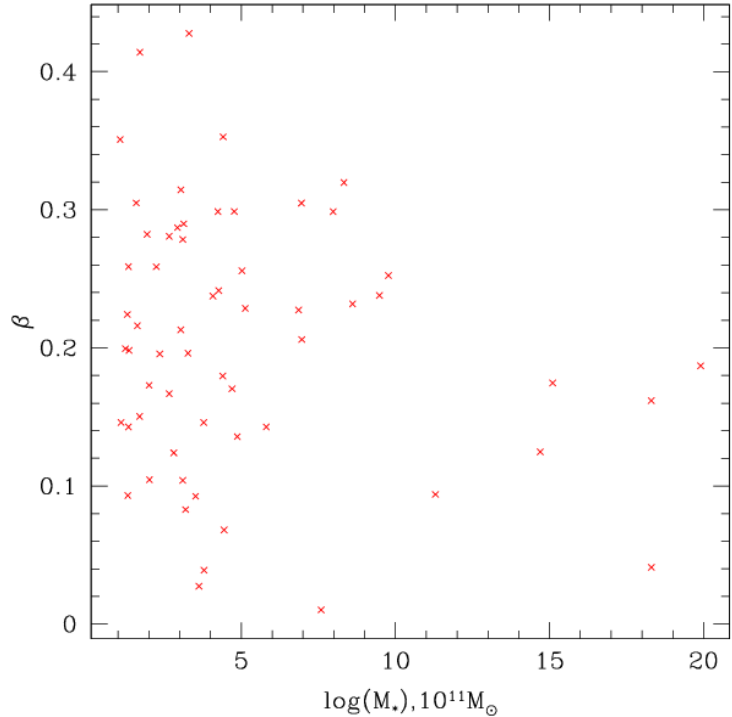

**Figure 5.** The dependence of the velocity dispersion anisotropy $\beta$ on the mass of sampled galaxies from the Illustris project.

As it can be seen, the E-galaxies with a very large mass have rather lower values of the velocity dispersion anisotropy $\beta$. Interestingly, this conclusion is consistent with our calculations: two of the most massive galaxies from our widen sample, NGC 1016 and NGC 315, also have relatively small magnitudes of anisotropy.

We drew attention to the fact that there is indeed an approximate coincidence of the velocity dispersions in one of the planes in 64% of the galaxies from Illustris, which confirms the correctness of this assumption used in our article. It should also be noted that only 25% of galaxies has a spatial shape of an oblate spheroid. Such a conclusion can be made using the data about their rotation and ellipticity, measured in three orthogonal projections. Most of the giant systems likely have either prolate, or triaxial shape, thus it confirms the need to use not only the models with rotational symmetry.

Systematic uncertainties of anisotropy magnitudes, obtained with our method, could be also estimated from comparison with Illustris project. But, unfortunately, we cannot perform such a fine analysis, since the data for $z = 0$ are not available for three-dimensional dispersions in Illustris galaxies.

## 6. DISCUSSION AND CONCLUSIONS

Despite a rapid progress in recent years, our understanding of the role of anisotropy of velocity dispersion in the global dynamics of elliptical galaxies remains not fully clarified. The solution of this problem requires: a) the multifold development of an adequate mathematical apparatus, capable of taking into account many features of the structure of E-galaxies; b) high-quality observational data, which could be fully used by the new methods. All this makes the task of construction models very difficult.

A key feature of our models is a rigorous consideration of the influence of the spatial shape (oblate spheroid or triaxial ellipsoid) on the dynamics of E-galaxies, as well as the structure of the internal density layers in these systems. Our method is based on the virial tensor theorem for an ellipsoidal layered inhomogeneous subsystem and allows us to take into account not only the flattening profile of isophotes and the spatial shape of E-galaxies, but also the change of the ratio $M/L$ from the center to the periphery of the galaxy (so far only for the M87 galaxy). Note that our models do not pretend to a full description of E-galaxies. The approach is based on a comparing of a simple dynamic isotropic model with real kinematic data of the galaxy and calculating the velocity dispersion anisotropy from this comparison. In other words, we find the value of the anisotropy of velocity dispersion as the degree of divergence between the isotropic model and observations.

This paper presents a pilot study to find the answer to the question of how the internal structure of E-galaxies affects their dynamics. When creating models, a huge amount of new observational information about elliptical galaxies was taken into account (the detailed surface brightness distributions, measured up to the far periphery and also rotation and dispersion curves). It is important that the question of the spatial shape of the model for each galaxy was resolved, taking into account triaxial tests. Our calculations were carried out for both axisymmetric oblate spheroid model and for a prolate spheroid, rotating end-to-end, dynamically equivalent to a triaxial ellipsoid.

The main task in our work was the calculation of the relation of the rotational energy to the module of the gravitational energy $t = \frac{T_{rot}}{|W|}$ for each galaxy, the ratio $\frac{\upsilon_{rot}}{\sigma}$, and the velocity dispersion anisotropy parameter $\beta$. On a sample of 12 E-galaxies, we found that the anisotropy parameter lies in the range

$$0.0 \leq \beta \leq 0.224, \qquad (42)$$

The analysis showed that the values of the anisotropy parameters of the velocity dispersion are significantly influenced by such factors as the spatial form of the models, the range of variation in the flattening of layers in the galaxy, and the specific type of rotation curves. The farther from the center of the galaxy these characteristics are measured, the more influence the structure of the layers has on the dynamics of E-galaxies.

Our results are carefully compared with the results of other researchers. It has been established that for small E-galaxies (fast rotators) our values $\beta$ are in satisfactory agreement with those, found in the $ATLAS^{3D}$ project, based on Jeans anisotropic models. However, for giant E-galaxies, our models, which take into account the shape and internal structure of these star systems from their center to the periphery, provide better agreement with observations than the axisymmetric models in this project.

We compared our results with the anisotropy of velocity dispersion $\beta$ of massive E-galaxies, obtained in high resolution cosmological modeling in the Illustris project. It is important that the anisotropy values, obtained from the velocity dispersions along the line of sight in three different projections of the modeled galaxies, turned out to be numerically quite close to our results. We see that the parameter $\beta$ for the most of galaxies lies within the interval $[0 \div 0.3]$, and only for some of them this parameter turns out to be slightly larger. Thus, most galaxies do have anisotropy $\beta$, like real systems in our study. It should also be noted that only 25% of giant Illustris galaxies have a shape of an oblate spheroid, which also confirms the need to use different models for analyzing the spatial shape of E-galaxies.

Comparison of our calculations with the results of high-resolution cosmological modeling performed in the Illustris project is a test of the adequacy of our calculations. In addition, such a comparison could be a criterion for verifying the viewpoint that the anisotropy of velocity dispersion is indeed an important factor in the formation and evolution of these stellar systems.

To study the evolution of galaxies, it is necessary to build their dynamic models. Summing up the results, it can be concluded that our virial approach made it possible to find out a subtle influence of the spatial shape and structure of layers of E-galaxies on their anisotropy of velocity dispersion. In the future, it is also important to find out how the anisotropy of velocity dispersion varies with distance from the center of a particular galaxy (see in connection with this Fig. 5). This will allow a more confident choice between different theories of the origin of E-galaxies. Additional effects, such as dark matter halo and isophote twisting, could be included in the model to increase its capability to recreate real galaxies. But, at this stage, our goal to modernize the virial method, making it suitable for studying the detailed structure of E-galaxies, may be considered as fulfilled.